\documentclass[twocolumn]{aastex631}

\usepackage[caption=false]{subfig}

\usepackage{amsmath}

\shorttitle{PSF reconstruction}
\shortauthors{Alonso et al.}

\graphicspath{{./}{figures/}}

\begin{document}

\title{A Hierarchical PSF Reconstruction Method}
\correspondingauthor{Jun Zhang}
\email{betajzhang@sjtu.edu.cn}

\author{Pedro Alonso}
\affiliation{Department of Astronomy, Shanghai Jiao Tong University, Shanghai 200240, China}
\affiliation{Shanghai Key Laboratory for Particle Physics and Cosmology, Shanghai 200240, China}

\author{Jun Zhang}
\affiliation{Department of Astronomy, Shanghai Jiao Tong University, Shanghai 200240, China}
\affiliation{Shanghai Key Laboratory for Particle Physics and Cosmology, Shanghai 200240, China}

\author{Cong Liu}
\affiliation{Department of Astronomy, Shanghai Jiao Tong University, Shanghai 200240, China}
\affiliation{Shanghai Key Laboratory for Particle Physics and Cosmology, Shanghai 200240, China}

\begin{abstract}
Reconstruction of the point spread function (PSF) plays an important role in many areas of astronomy, including photometry, astrometry, galaxy morphology, and shear measurement. The atmospheric and instrumental effects are the two main contributors to the PSF, both of which may exhibit complex spatial features. Current PSF reconstruction schemes typically rely on individual exposures, and its ability of reproducing the complicated features of the PSF distribution is therefore limited by the number of stars. Interestingly, in conventional methods, after stacking the model residuals of the PSF ellipticities and (relative) sizes from a large number of exposures, one can often observe some stable and nontrivial spatial patterns on the entire focal plane, which could be quite detrimental to, e.g., weak lensing measurements. These PSF residual patterns are caused by instrumental effects as they consistently appear in different exposures. Taking this as an advantage, we propose a multi-layer PSF reconstruction method to remove such PSF residuals, the second and third layers of which make use of all available exposures together. We test our method on the i-band data of the second release of Hyper Suprime-Cam. Our method successfully eliminates most of the PSF residuals. Using the Fourier\_Quad shear measurement method, we further test the performance of the resulting PSF fields on shear recovery using the field distortion effect. The PSF residuals have strong correlations with the shear residuals, and our new multi-layer PSF reconstruction method can remove most of such systematic errors related to PSF, leading to much smaller shear biases.
\end{abstract}


\section{Introduction} \label{section:Introduction}
The point spread function (PSF) measures the diffraction of light in optical systems, which makes point-like sources appear extended. In astronomical images, the PSF effect distorts the shape and size of all celestial objects. For ground-based telescopes, the properties of the PSF are mainly determined by the telescope optics \citep{Jarvis_2008} as well as the atmospheric turbulence \citep{1981PrOpt..19..281R, 2018MNRAS.481.5210F, 2018SPIE10700E..5EH, 2018AJ....156..222X}. A minor contributor to the PSF is the pixelation effect, resulting from the finite pixel size of the CCD images \citep{2007PASP..119.1295H, 2010MNRAS.403..673Z,Kannawadi_2021, 2022AJ....164..214S,Hirata_2024}.

PSF plays a particularly important role in weak lensing studies, since it convolves the lensed galaxies and distorts their shapes akin to the cosmic shear. This makes the PSF effect the most important source of systematics in shear measurements \citep{2007ApJS..172..203R, 2013MNRAS.429.2858M, 2017AJ....153..197L, 2023RAA....23g5021L}. To ensure the reliability of the measurements, precise modeling of the PSF is essential.

When modeling the PSF, it is convenient to describe it as a combination of time-variant and time-invariant features. Time-variant features vary from exposure to exposure and are produced by the atmospheric turbulence and some instrument-related issues such as misalignments in the optical components of the telescope, mechanical deformations caused by gravitational or thermal effects, tracking errors, or instrumental instabilities. On the other hand, time-invariant features are quite stable from exposure to exposure.
Optical designs, mechanical elements of the telescope, as well as characteristics of the imaging sensor, focal length, or aperture size can all induce time-invariant features on the PSF.

Multiple methods have been proposed to model the PSF effect for shear measurement \citep{1995ApJ...449..460K,1997ApJ...475...20L,1998ApJ...504..636H,2000ApJ...537..555K,2014MNRAS.438.1880B,2015JCAP...01..024Z,2016MNRAS.459.4467B}. With time-variant features complicating the PSF modeling, most of these methods rely solely on individual exposures to build the PSF model, thus being limited by the number of available stars. An effective way to deal with a limited number of stars is to assume a specific functional form for the distribution of the PSF, e.g., polynomial functions. Several methods that follow this approach have been successful in capturing most of the PSF features on the exposure level. Other methods, based on Principal Component Analysis (PCA) \citep{Jarvis_2004} or machine learning \citep{2018JCAP...07..054H, 2020AJ....159..183J}, have also shown impressive results in capturing the spatial features of the PSF. The optimal PSF reconstruction method generally depends on the particular shear measurement being considered, and the particular dataset.

In current methods, however, when stacking the PSF ellipticity residuals of the PSF models of many exposures, some complex features appear \citep{Bosch_2017,Jarvis_2020,Zhang_2022}. These features are time-invariant, and therefore related to instrumentation. In this work, we present a novel PSF reconstruction method that performs three interpolations in a hierarchical manner, with the aim of modelling the high-frequency yet stable spatial features of the PSF distribution. Our method builds upon the PSF reconstruction method of Liu et al. 2024 (in preparation), constructing a model of the PSF power spectrum. We apply our method to the i-band of HSC pDR2 data, in which systematic PSF residuals have been previously reported. Our method successfully removes most of the systematic PSF residuals found in HSC DR2, significantly improving the PSF model.

The structure of this paper is as follows. \S\ref{section:HSC_data} introduces the HSC dataset. The first layer of interpolation is explained in \S\ref{section:layer1}. The second and third layers of interpolation are described in \S\ref{section:layer2}, which also includes our main results. We conclude in \S\ref{section:Conclusion}.

\begin{figure}[htb]
    \begin{center}
    \includegraphics[width=0.47\textwidth]{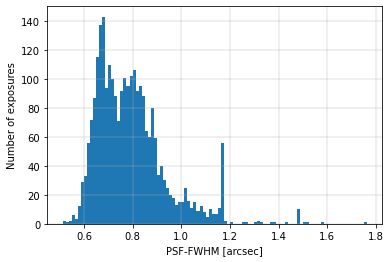}
	\caption{Distribution of the PSF FWHM of the i-band of HCS pDR2.}
	\label{fig:psf_fwhm_distrib}
    \end{center}
\end{figure}

\section{HSC dataset}
\label{section:HSC_data}
The Hyper Suprime-Cam Subaru Strategic Program (HSC-SSP) project \citep{2018PASJ...70S...4A,2018PASJ...70S...1M,2018PASJ...70S...2K,2018PASJ...70S...3F}
is an optical multi-layer imaging survey that covers approximately $1400~\mathrm{deg}^2$ in five bands ($g, r, i, z, y$) in its \textit{Wide} layer ($r\sim26$). In addition, the survey includes two deeper layers, \textit{Deep} and \textit{UltraDeep}, covering $27~\mathrm{deg}^2$ ($r\sim27$) and $3.5~\mathrm{deg}^2$ ($r\sim28$), respectively. The project utilizes the Hyper Suprime-Cam, a wide field optical camera built on the 8.2m Subaru Telescope, and aims to address some of the most important problems in astrophysics and comsmology, with a focus on weak gravitational lensing, galaxy evolution, supernovae, and galactic structure. Their data is publicly available at their official website\footnote{https://hsc-release.mtk.nao.ac.jp/}.

In this paper we use the i-band data of the second public data release of the HSC \citep[HSC pDR2;][]{2019PASJ...71..114A}. HSC pDR2 covers an area of 300 square degrees in the \textit{Wide} layer, in all five bands. The data was collected over 174 nights of observation, from March of 2014 to January of 2018. HSC pDR2 includes significant improvements over the previous release (HSC pDR1), which include improved background subtraction, PSF modeling, and object detection procedures. However, we only use the background-removed CCD images, performing our own PSF reconstruction. The galaxies are selected from the official HSC catalog \citep{2019PASJ...71..114A}. Fig.~\ref{fig:psf_fwhm_distrib} shows the distribution of the PSF full width at half maximum (FWHM), in real space, for the $i$ band of HSC pDR2.

\section{PSF reconstruction}
\label{section:layer1}

\begin{figure*}[htb]
    \begin{center}
        \includegraphics[width=\textwidth]{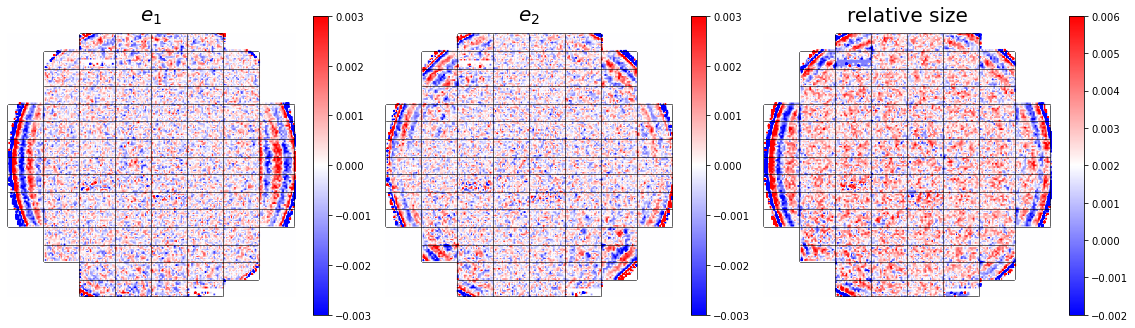}
	\caption{Stacked ellipticity ($e_1$, $e_2$) and relative size residuals from all exposures of the i-band of HCS pDR2.}
	\label{fig:residuals}
    \end{center}
\end{figure*}

Our method builds upon the PSF reconstruction method of Liu et al. 2024 (in preparation), which is part of the Fourier\_Quad shear measurement pipeline \citep[]{Zhang_2022}. Fourier\_Quad utilizes the quadrupole moments of the power spectrum of galaxy images to measure the cosmic shear. In line with this approach, we perform PSF interpolation directly on the power spectrum of the stars, building a model of the power spectrum of the PSF (PPS hereafter). 

\subsection{Star selection}
Several techniques have been developed for star-galaxy separation. Some methods rely on morphological features \citep{2020AJ....159...65S} or color information of the sources \citep{2010A&A...514A...3P}. Moreover, modern approaches based on machine learning methods such as decision trees \citep{2011AJ....141..189V} or deep convolutional neural networks \citep{2017MNRAS.464.4463K} have shown promising results in star-galaxy separation.

In this work, we follow the following procedures to select out bright stars for PSF reconstruction (Liu et al. 2024, in preparation):

1. For each exposure, we select sources with ${\rm SNR}\geq100$ to form our initial set of star candidates. The rest of the steps only make use of their power spectra, which are all normalized so that the power $P(\vec{k}=0)$ is unity. Note that in Fourier space, point sources should have the most extended profiles. We therefore measure the area of each candidate (defined as the number of pixels above 0.02). The distribution of the area typically has a Gaussian shape. We throw away those candidates that are away from the peak of the distribution by more than $3\sigma$ to remove the outliers. 

2. From the remaining star candidates, we form the first star model (power spectrum) pixel-by-pixel.
Each pixel value is determined by sorting the corresponding pixel values from the power of all candidates, and taking the lower bound of the top 25\%.

3. The similarity between the candidates and the star model can be quantified by defining a $\chi^2$ as:
\begin{equation}
\label{c2}
\chi^2=2\sum_{i=1}^{N}\left(I_i^n-I_i^{model}\right)^2/\sum_{i=1}^{N}\left(I_i^n+I_i^{model}\right)
\end{equation}
in which $I_i^n$ refers to the value of the $i^\text{th}$ pixel in the power image of the $n^\text{th}$ candidate, and $I_i^\text{model}$ refers to the corresponding pixel value of the model. $N$ is the total number of pixels involved, which are typically chosen to be those located in the middle part of the stamp. The distribution of the $\chi^2$ forms a Gaussian-like function, and we remove those candidates whose $\chi^2$ are more than $3\sigma$ away from the peak of the distribution. 

4. We build the new star model, this time as a function of location, by interpolating the pixel values of all the remaining candidates with polynomial functions of order nine. Note that if there are not enough candidates left for the fitting, we simply stop processing the exposure further. We again use eq.(\ref{c2}) to define the $\chi^2$ between each candidate and the star model at its location. From the distribution of $\chi^2$, we again remove the candidates that are more than $3\sigma$ away from the peak. The surviving candidates are treated as stars for our PSF reconstruction. 

\subsection{First layer of interpolation}
\label{sec:1st_interp}
Our first interpolation is a 2D polynomial fitting of third order on the power spectrum of the stars on each CCD image, following the procedure of Liu et al. 2024 (in preparation). To ensure the reliability of the PSF model, we only include CCDs containing a minimum of 20 stars. Each star power spectrum is centered on a $48$x$48$ stamp, and the interpolation is done pixel-by-pixel.

To evaluate the quality of the interpolation, we calculate the PPS residuals, represented as the ellipticity and relative size residuals at the positions of the stars. The ellipticity components, $e_1$ and $e_2$, and the size are defined based on the quadrupole moments, $Q_{ij}$, as:

\begin{equation}
\begin{aligned}
    e_1 =& \frac{Q_{20}-Q_{02}}{Q_{20}+Q_{02}}\\\\
    e_2 =& \frac{2Q_{11}}{Q_{20}+Q_{02}}\\\\
    size = &\frac{Q_{20}+Q_{02}}{Q_{00}},\\
\end{aligned}
\end{equation}
where
\begin{equation}
    \label{eq_Qij}
    Q_{ij} = \sum_{P(k)>P_0} P(k)k^i_x k^j_y.
\end{equation}
We only include pixels above $P_0=0.02\cdot P(k=0)$ in the calculation.

Ellipticity residuals are calculated as the difference between the ellipticities of the original ($e_{1,true}$, $e_{2,true}$) and the predicted ($e_{1,pred}$, $e_{2,pred}$) stamps, while the relative size residuals are calculated as $(size_{true}-size_{pred})/size_{true}$. Fig.~\ref{fig:residuals} shows the stacked ellipticity and relative size residuals including all exposures. While the residuals are very small for most part of the exposure, we find some remaining systematic residuals near the boundaries, which are not visible on single exposures but become prominent after stacking a large number of them. Note that similar PSF residual patterns are also observed with the official HSC pipeline \citep{Bosch_2017}, as well as in the DES data \citep{Jarvis_2020} and the DECaLS data \citep{Zhang_2022}. In the next layer of interpolation, we build a model for these systematic PSF residuals.

\section{Improved PSF reconstruction}
\label{section:layer2}
\begin{figure*}[htb]
    \begin{center}
	\includegraphics[width=\textwidth]{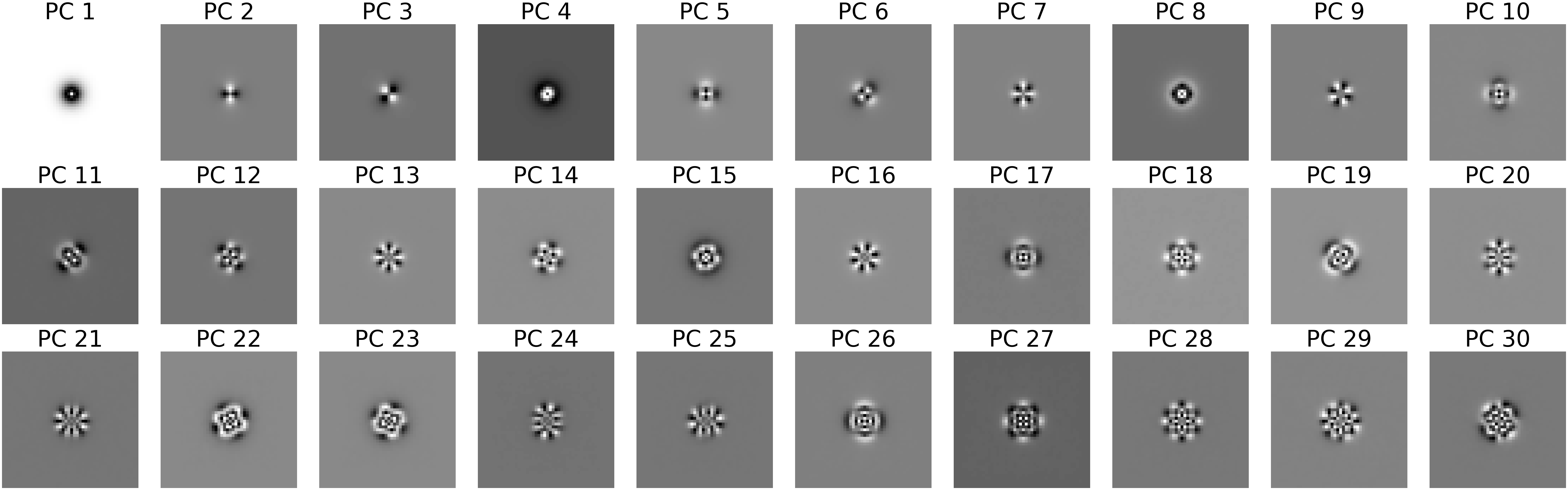}
	\caption{First 30 principal components of the CCD number 50, located in the central region of the exposure.}
	\label{fig:pcs_ccd50}
    \end{center}
\end{figure*}

\begin{figure*}[htb]
    \begin{center}
	\includegraphics[width=\textwidth]{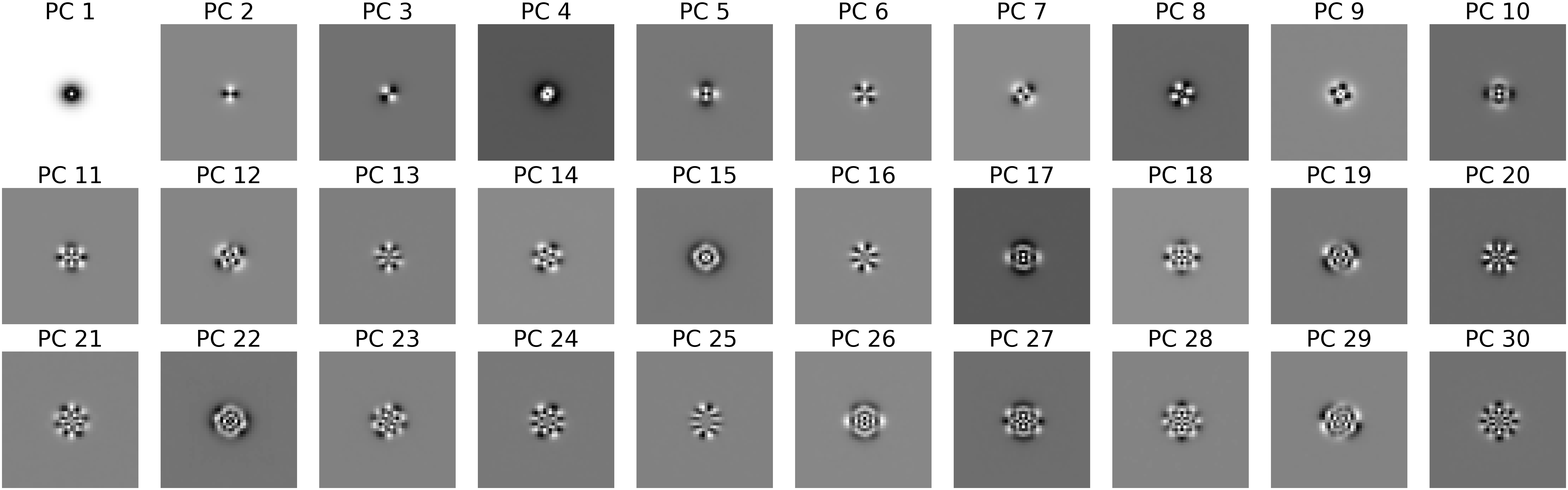}
	\caption{First 30 principal components of the CCD number 53, located in the boundary of the exposure.}
	\label{fig:pcs_ccd53}
    \end{center}
\end{figure*}

\begin{figure*}[htb]
    \begin{center}
	\includegraphics[width=\textwidth]{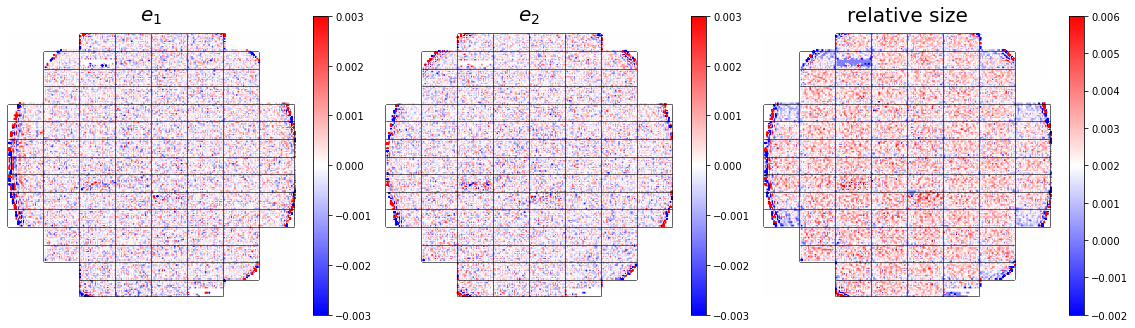}
	\caption{Stacked ellipticity ($e_1$, $e_2$) and relative size residuals of all exposures of the i-band of HSC pDR2, after the second interpolation, for the case of a polynomial fitting of order 6 as our second interpolation.}
	\label{fig:residuals_corr}
    \end{center}
\end{figure*}

\begin{figure*}[htb]
    \begin{center}
	\includegraphics[width=\textwidth]{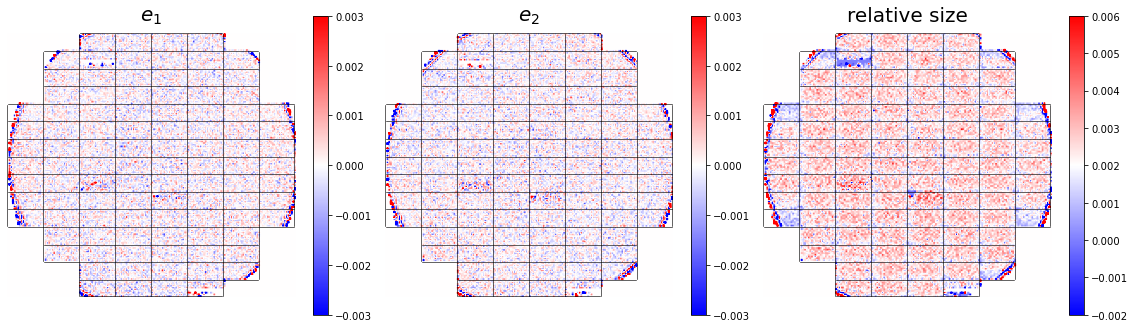}
	\caption{Stacked ellipticity ($e_1$, $e_2$) and relative size residuals of all exposures of the i-band of HSC pDR2, after the second interpolation, for the case of random forest as our second interpolation.}
	\label{fig:residuals_corr_rf}
    \end{center}
\end{figure*}

The systematic PPS residuals observed in fig.~\ref{fig:residuals} are related to instrumental effects rather than the atmospheric turbulence. Given the strong spatial and temporal dependencies of the atmospheric turbulence, it is unlikely to produce any persistent features on the PSF residuals.
In the following we present our second layer of interpolation, which builds a model for the systematic features of the PPS residuals. Unlike in the first interpolation, the features that we want to model are systematic, common to all exposures. Therefore, instead of performing an interpolation on individual chips, we collect the PPS residual stamps from all exposures and place them into a single exposure. We then perform a single interpolation on each CCD of that exposure. This procedure significantly increases the amount of available data for interpolation.

To collect the PPS residuals from all exposures and place them into a single exposure, we must first take into account the different PSF sizes in different exposures (fig.~\ref{fig:psf_fwhm_distrib}). We rescale the PPS residual stamps to a common size, with the re-scaling factor determined as the mean PSF size of the exposure. It is important to emphasize that the above procedure of rescaling the PSF stamps from different exposures and place them into a single exposure is only beneficial when modelling systematic features, common to all exposures. In our first interpolation, the main contributor to the PSF was the atmospheric turbulence, which strongly varies from exposure to exposure, hence the interpolation was performed on individual exposures.
 
\subsection{Rescaling of the PPS residuals}
\label{sec:rescaling}

\begin{figure*}[htb]
    \begin{center}
	\includegraphics[width=\textwidth]{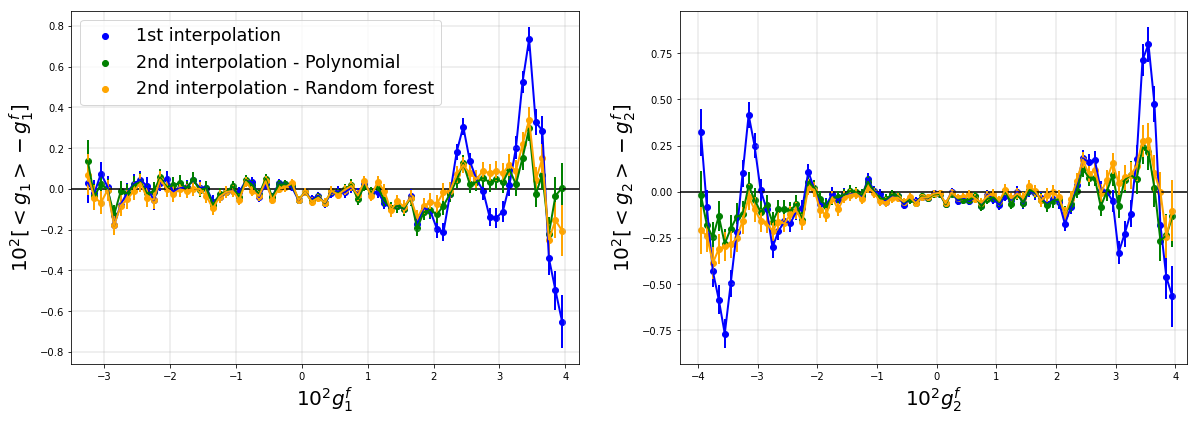}
	\caption{Field distortion test after first interpolation (blue curves), and after the second interpolation, for the case of polynomial of order 6 (green curves) and random forest (yellow curves). Results are shown for $g_1$ (left panel) and $g_2$ (right panel). Data points show $1-\sigma$ errorbars.}
	\label{fig:FDS_compare}
    \end{center}
\end{figure*}

We rescale each PPS residual to a reference PSF FWHM of $1$ arcsec in real space. For each PPS residual stamp, we calculate the rescaling factor $s$ as the mean PSF FWHM size, in real space, of all stars in the same CCD in unit of arcsec. To build the rescaled PPS residual stamp, we project the coordinates $(u, v)$ of each pixel back to the original stamp, as $(u\cdot s, v\cdot s)$. In most cases, the projected coordinates $(u\cdot s, v\cdot s)$ lies within four pixels of the original stamp. The value of the $(u, v)$ coordinate in the rescaled stamp is calculated as a weighted average of those four neighboring pixels in the original stamp, with the weights determined by the inverse of the pixel distances to $(u\cdot s, v\cdot s)$.

\subsection{Principal Component Analysis (PCA) of the PPS residuals}
\label{sec:PCA}

Once the PPS residual stamps have been rescaled, we place them into a single exposure, and apply principal component analysis to the stamps on each CCD. Principal component analysis \citep[PCA;][]{2014arXiv1404.1100S} is a widely used method for dimensionality reduction in data analysis. It transforms an N-dimensional system into a lower dimensional representation, characterized by Principal Components, or PCs. These PCs form the basis of the new space, representing the main features of the data. They are orthogonal to each other, which helps eliminate the correlation between variables, making PCA particularly useful for large datasets, where numerous correlated variables make data interpretation difficult. The main motivation for applying PCA to the PPS residuals is to capture the main features of the data, eliminating unnecessary information or noise. Additionally, dimensionality reduction drastically decreases the computational time, since instead of performing the interpolation pixel-by-pixel, now we perform an interpolation for each principal component.

We choose 100 as the number of principal components, resulting in each stamp being represented by the 100 coefficients to the PCs. Fig.~\ref{fig:pcs_ccd50} and fig.~\ref{fig:pcs_ccd53} show the first 30 principal components of the CCDs number 50 and 53, respectively \citep{2018PASJ...70S...4A}. The CCD number 50 is located in the central region of the exposure, whereas the CCD number 53 is located in the boundary. We observe that their first four principal components are almost identical, diverging from the fifth, showing the different features of the PPS residuals in the two regions.

\subsection{Reconstruction of the PPS residuals}
\label{sec:rec_residuals}

We build a model for each principal component coefficient at the CCD level by interpolating the PC coefficients of the PPS residuals within that CCD. We test two different interpolation methods, polynomial and random forest, and compare their performance.

\subsubsection{Polynomial}
\label{sec:rec_pol}

The features of the PPS residuals are particularly complex near the boundaries of the exposure, thus a single interpolation on a CCD level is unlikely to fully capture them. Since the amount of data per CCD is now very large, we further divide each CCD into four equal parts ($2\times 2$ in the CCD plane), and fit a polynomial of order six to the PC coefficients of the PPS residuals within each part. The predicted PPS residual stamp ($48$x$48$) at the position of each star is constructed as the sum of the PCs, weighted by the predicted PC coefficients. Finally, we rescale the current PPS residuals model back to the original PSF size at each position. We follow the same procedure as \S\ref{sec:rescaling}, with the scaling factor being the inverse of $s$.

The improved PPS model is built by adding the PPS residuals model to the original PPS model from the first interpolation. Fig.~\ref{fig:residuals_corr} shows the new ellipticity and relative size residuals, calculated at the position of the stars, as the difference between the ellipticity components/relative size of the true star and the new PPS model. We observe a significant improvement, with the systematic PPS residuals vanishing almost completely.

\subsubsection{Random Forest}
\label{sec:random_forest}
In this section we follow the same procedure as in \S\ref{sec:rec_pol}, but using a machine learning algorithm called random forest to interpolate the principal component coefficients of the PPS residuals. Random Forest \citep{598994, Breiman2001RandomF} is a widely used machine learning algorithm, applicable to both classification and regression problems. It is an ensemble method that combines the predictions of multiple models, called decision trees, to infer a final prediction. Each decision tree is built on a randomly selected subset of the data, and the output of the random forest is the average output of all the decision trees. This significantly reduces the risk of overfitting.

Each decision tree is built through a series of binary splits of the data, starting from the root node. At first, two subnodes of the root node are created, based on a splitting condition. Following the same procedure, each of the subnodes is further divided into two new subnodes, based on new splitting conditions, and so on, building the tree. When a stopping condition is met, we stop splitting that node.

In our case, each decision tree performs a regression on a randomly selected subset of our data, where each datapoint is described by the coordinates ($x, y$) on the CCD, and the principal component coefficient value, $z$. Starting from the root node, we perform the first split of the data. The algorithm splits the data into two groups or subnodes (A and B), based on a condition\textemdash either on $x$ or $y$\textemdash that better splits the data according to the values of $z$. It is important to highlight that the algorithm sees $x$ and $y$ as distinct features of the data, rather than as coordinates. As a result, each split is based on a single feature, either $x$ or $y$. To determine the optimal splitting condition for each node, we use the mean squared error (MSE), a widely adopted metric for splitting in regression tasks. We define the MSE of each possible split as:
\begin{equation}
    MSE = \sum_{(x,y)\in A}(z-z_{pred, A})^2+\sum_{(x,y)\in B}(z-z_{pred, B})^2,
\end{equation}
where $z_{pred, A}$ and $z_{pred, B}$ are the predicted PC coefficients associated to each group, and it is the same for all elements in the group. The split with the lowest MSE value is the optimal split.

We continue splitting each subnode (A and B), following the same procedure as for the root node. We stop splitting a node when it contains less than 500 datapoints. This stopping criterion is found empirically, as further splitting the data results in overfitting.

A decision tree generates predictions for new coordinates ($x, y$) by following the splitting conditions for each node. Starting at the root node and moving down the tree, the new data reaches a leaf node, i.e., a node without subnodes. The predicted PC coefficient for the new coordinates is the value associated to that leaf node, $z_{pred}$.

As in \S\ref{sec:rec_pol}, we divide each CCD into four parts ($2\times 2$ in the CCD plane), and build models for the PC coefficients of the PPS residual stamps within each part. We build 100 decision trees for each principal component coefficient, and determine the predicted PC coefficient at each CCD position, $z_{pred}$, as the average predictions of all decision trees. We build the PPS residual model as the weighted sum of the principal components, with the weights given by the PC coefficients predicted by random forest, $z_{pred}$. To obtain the final PPS residual model at each CCD position, we rescale the PPS residual model back to the PSF size of each exposure. Fig.~\ref{fig:residuals_corr_rf} shows the ellipticity and relative size residuals of our improved PPS model, calculated as the sum of the PPS model from the first interpolation and the PPS residual model. As in the polynomial case, we obtain very small PPS residuals, removing almost completely the systematic PPS residuals of fig.~\ref{fig:residuals}. Although polynomial and random forest obtain comparable results, the PPS residuals of random forest are slightly smaller. However, this does not necessarily mean that random forest is going to be more accurate in making predictions at the position of galaxies. As a non-parametric algorithm, random forest could in principle capture more complex features on the data. However, they are also more prone to overfitting. To evaluate the predictions at galaxy positions of our new PPS models, we use the field distortion test.

\subsection{Field distortion test}
\label{sec:FD_test}
The field distortion \citep[FD;][]{2019ApJ...875...48Z} is an optical aberration characterized by the deviation from global rectilinear projection. It induces a distortion in the shape of galaxies in a similar way as the cosmic shear, which can be directly derived from astrometry parameters. \cite{2019ApJ...875...48Z} proposed a method to use the field distortion to evaluate the accuracy of shear recovery directly on real galaxies, by comparing the measured field distortion induced shear (FDS) and the true field distortion inferred from astrometry. This is known as the field distortion test. We refer to \cite{2019ApJ...875...48Z} for a derivation of the FDS equations.

We use the FD test to evaluate the performance of our PSF model on real galaxies, and compare the results of polynomial and random forest. We evaluate the true FDS, $g_1(\mathrm{FD})$ and $g_2(\mathrm{FD})$, against the FDS recovered from galaxies, $g_1(\mathrm{gal})$ and $g_2(\mathrm{gal})$.

Fig.~\ref{fig:FDS_compare} compares the FD test results for $\mathrm{SNR_F}\geq 4$ \citep{2021ApJ...911..115L}, after the first and second interpolation. Note that the FDS signals are removed from the galaxies. The x-axis represents the true FD signal and the y-axis the difference between the recovered FD signal from astrometry and the true one. The results after the first interpolation clearly show the imprints of the PPS residuals of fig.~\ref{fig:residuals}, and suggest that all galaxies with a FD signal of $|g_{1,2}(\mathrm{FD})|>0.02$ should be removed from the shear catalogs, as their PSF models are not reliable. We find a significant improvement after the second interpolation, particularly in the $|g_{1,2}(\mathrm{FD})|>0.02$ range. This improvement allows the shear catalog to include a much larger number of galaxies, thus enhancing its statistical power.

Fig.~\ref{fig:FDS_compare} shows that although random forest obtained slightly smaller PPS residuals compared to the polynomial case, both methods perform equally well in predicting the PPS model at galaxy positions. This results demonstrate the efficacy of the FD test in evaluating the PSF reconstruction accuracy directly on galaxy positions. 

\subsection{Further improvements}
\label{sec:3rd_interp}

\begin{figure}[htb]
    \begin{center}
	\includegraphics[width=0.46\textwidth]{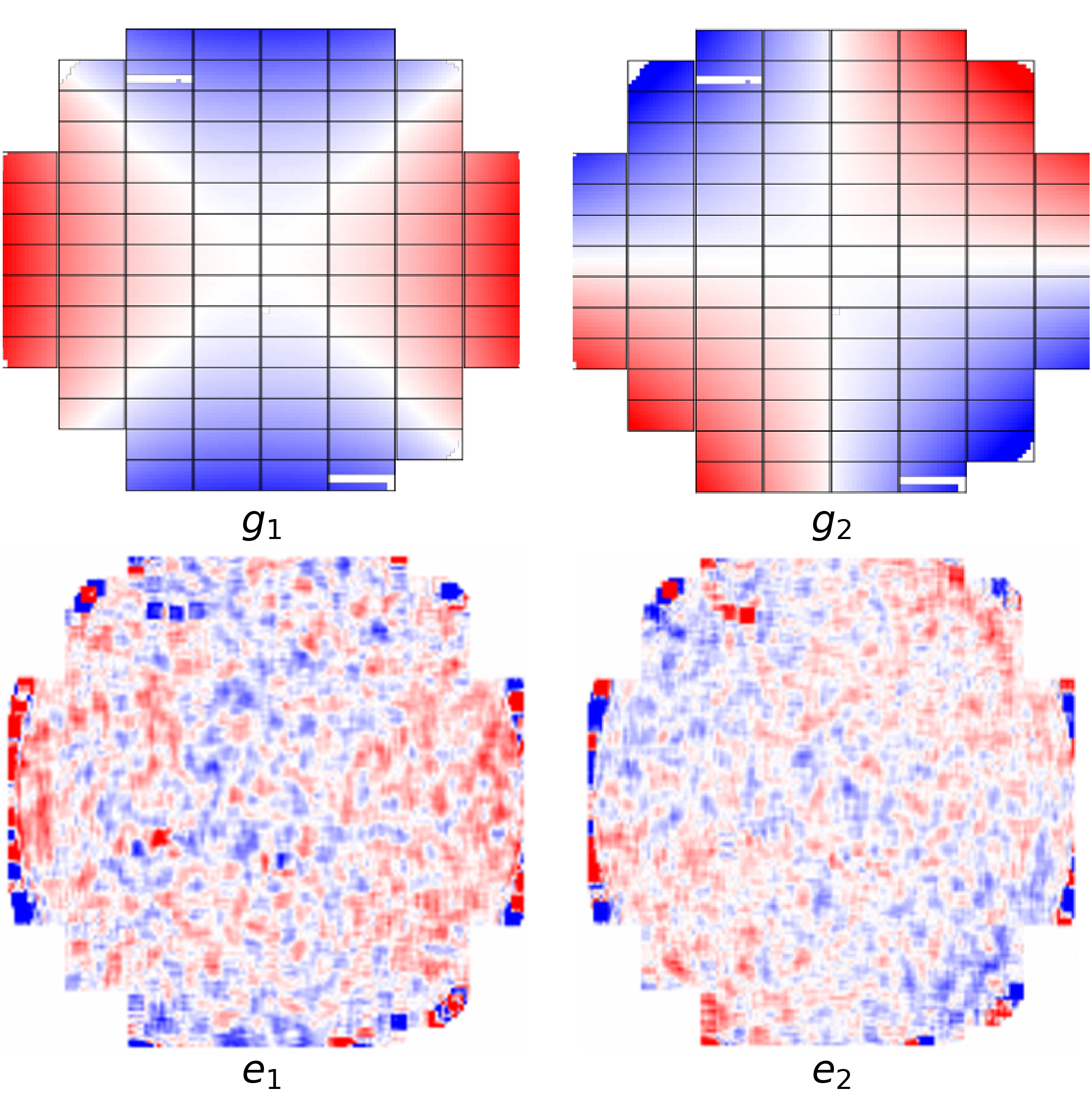}
	\caption{Field distortion signals ($g_1$, and $g_2$, top panels) and smoothed PPS residuals ($e_1$, and $e_2$, bottom panels).}
	\label{fig:res_vs_fd_HSC}
    \end{center}
\end{figure}

\begin{figure*}[htb]
    \begin{center}
        \includegraphics[width=\textwidth]{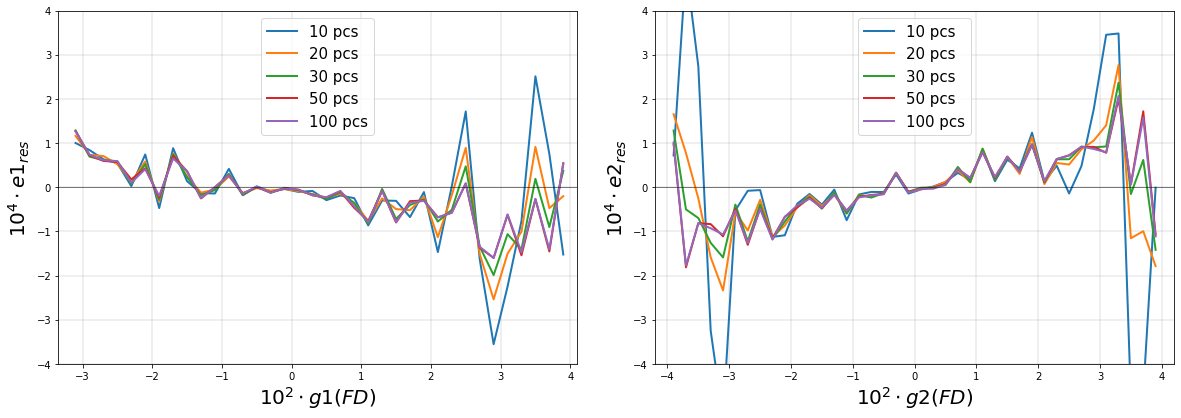}
	\caption{PPS residuals versus field distortion signal at the position of the stars. Different curves represent different number of PCs. The left panel shows the results for $e_1$ vs $g_1$ and the right panel shows the results for $e_2$ vs $g_2$.}
	\label{fig:npcs}
    \end{center}
\end{figure*}

\begin{figure}[htb]
    \begin{center}
	\includegraphics[width=0.47\textwidth]{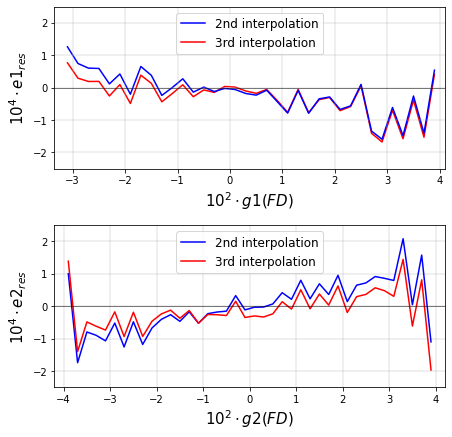}
	\caption{PPS residuals versus field distortion signal at the position of the stars. We compare the results after the second interpolation (blue curves) and after the third interpolation (red curves). The top panel shows the results for $e_1$ vs $g_1$ and the bottom panel shows the results for $e_2$ vs $g_2$.}
	\label{fig:res_vs_fd_corr}
    \end{center}
\end{figure}

After smoothing the PPS residuals shown fig.\ref{fig:residuals_corr}, we find an interesting phenomenon: the PPS residuals show a very small but systematic slope, similar to that shown by the field distortion induced shear (FDS). Fig.\ref{fig:res_vs_fd_HSC} displays the FDS ($g_1$ and $g_2$), and the ellipticity residuals ($e_1$ and $e_2$), which show a clear correlation. This correlation was not observed earlier mainly because this effect is too small and it was covered by the noisy PPS residuals signal.

We perform a third interpolation that aims to correct for this correlation. As we observe in fig.\ref{fig:res_vs_fd_HSC}, the systematic PPS residual is now global, thus we perform a single interpolation on the scale of the whole exposure.

As in the second layer of interpolation, we first rescale the PPS residual stamps\textemdash obtained after the second interpolation\textemdash to a reference PSF FWHM of $1$ arcsec, in real space. We place them into a single exposure and apply principal component analysis (\S\ref{sec:PCA}) to the PPS residual stamps in entire exposure. We use 100 principal components in our analysis. To find this number, we evaluate the quality of our PPS model for different number of PCs, using the field distortion test. Fig.~\ref{fig:npcs} shows the correlation between the PPS residuals ($e_1$ and $e_2$) and the FD signal ($g_1$ and $g_2$) for different number of PCs (10, 20, 30, 50, and 100). From 10 to 50 PCs, we observe an clear improvement as the number of PCs increase, with the results stabilizing after 50 PCs, without a noticeable improvement. Based on these results, we conclude that a minimum of 50 PCs is necessary to capture all the features of the PPS residuals. However, it is important to note that these results might vary depending on the specific dataset, thus we recommend maintaining the number of PCs above 50, ideally 100, to ensure optimal performance.

Using all available data, each set of PC coefficients is fit to a polynomial of third order, building a model for the PC coefficient. These models are used to predict the PC coefficients at different positions on the exposure. Next, a model for the PPS residuals is built for each coordinate on the exposure as a weight sum of the PCs, weighted by the predicted PC coefficient at that position. Lastly, we rescale the PPS residual model back to its original PSF size, as in \S\ref{sec:rec_residuals}. This PPS residuals model is added to the PPS model after the second interpolation, building our final PPS model.

Fig.~\ref{fig:res_vs_fd_corr} compares the PPS residuals with the field distortion shears after the second and third interpolations. Although we observe a slight improvement after the third interpolation, this is not quite significant. We have explicitly tested various polynomial orders for our third interpolation, ranging from 2 to 6. The third order demonstrated the best performance. In addition, we have also tested our interpolation using random forest, but in this case it performs significantly worse than polynomial, likely due to overfitting.

\begin{figure*}[htb]
    \begin{center}
	\includegraphics[width=0.9\textwidth]{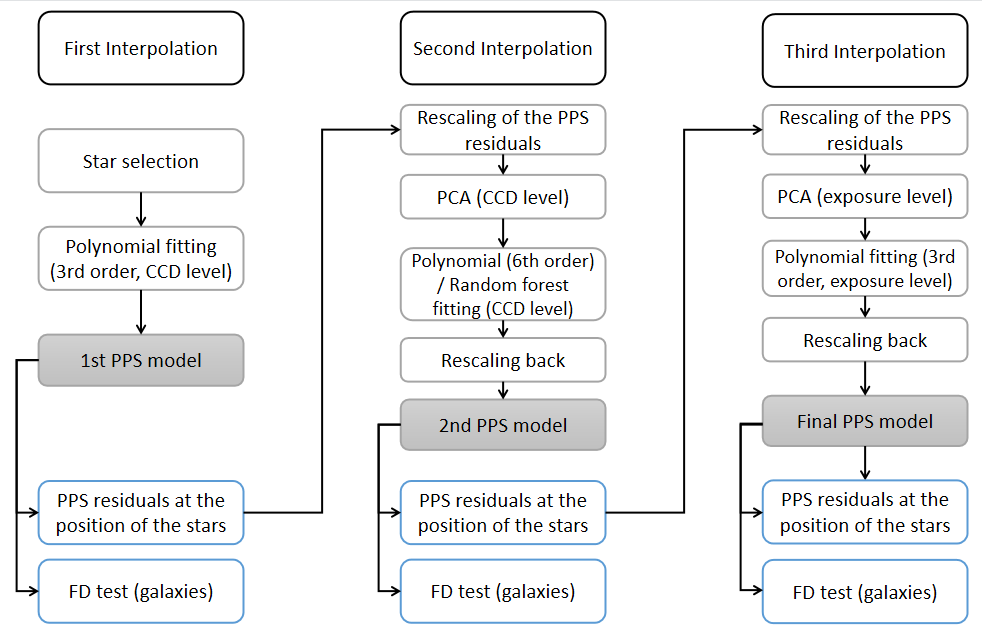}
	\caption{Structure of our PSF reconstruction method, including three layers of interpolation.}
	\label{fig:flow_chart}
    \end{center}
\end{figure*}

We summarize all the steps of our PSF reconstruction scheme in a flowchart shown in fig.~\ref{fig:flow_chart}.

\section{Conclusion} \label{section:Conclusion}
Point spread function (PSF) reconstruction is a crucial step towards accurate shear measurements, as it directly affects the observed galaxy morphology. Although current PSF reconstruction methods can capture most of the PSF features, there are often some PSF residuals remaining that limit the accuracy of, e.g., shear measurements. It is therefore essential to develop new techniques that can further improve current PSF reconstruction methods. In this work we introduce a novel PSF reconstruction method composed by three layers of interpolation, following a hierarchical scheme. Using data from the second data release of the Hyper Suprime-Cam (HSC DR2) as an example, our method significantly reduces the systematic PSF residuals, improving the PSF model.

The first layer of interpolation (\S\ref{sec:1st_interp}) is a 2D polynomial fitting of third order to the power spectrum of the stars, pixel-by-pixel, on each CCD separately. This interpolation captures most of the spatial features of the PSF, building our initial PSF model.

The second layer of interpolation, described in \S\ref{section:layer2}, is the most important and novel part of this work. In this layer we model the systematic PSF residuals remaining after the first interpolation, using the data of all the exposures simultaneously. Note that to do so, we need to first uniform the image sizes of the PSF residuals. We apply PCA to the PSF residuals and build a model for each principal component coefficient. We then build the PSF residual model as a weighted sum of the principal components, with the weights being the predicted PC coefficients. To make corrections on the PSF model, we rescale the PSF residuals model back to the original PSF size of each exposure and add it to the original PSF model. Fig.~\ref{fig:residuals_corr} and fig.~\ref{fig:residuals_corr_rf} show the PSF residuals of our improved PSF model, for polynomial and random forest interpolations, respectively. In both cases our method successfully removes most of the systematic features of the PSF residuals. In addition, we test our model directly on real galaxies, using the field distortion test (\S\ref{sec:FD_test}). We obtain significant improvements compared to the results of the first interpolation. Fig.~\ref{fig:FDS_compare} shows the field distortion test results for the cases of polynomial and random forest interpolations. Both interpolations have comparable performances, leading to very small shear biases.

In \S\ref{sec:3rd_interp} we study a correlation between the PSF residuals and the field distortion signal, which is found after smoothing the PPS residuals after the second interpolation (see fig.\ref{fig:res_vs_fd_HSC}). In this case, the systematic features are global, hence we perform a single interpolation on the exposure level, making use of all available exposures together. As in the second interpolation, we first rescale the PSF residuals to a common size. Then, we apply PCA to the entire data, and build a model for each PC coefficient. Our new PSF residuals model is built as a weighted sum of the principal components, with the weights given by the predicted PC coefficient at each position on the CCD. Lastly, we rescale the PSF residuals model back to the original PSF size of each exposure, and add it to the PSF model. Fig.\ref{fig:res_vs_fd_corr} presents the PPS residuals versus the field distortion shears, showing a small improvement after our third interpolation, although not quite significant. Nevertheless, this effect is minor and does not significantly impact the PSF model.

Overall, our model introduces a way to model the systematic PSF residuals, successfully removing most of the systematic PSF residuals in HSC DR2. In the current framework, our machine learning approach (random forest) performs similarly to polynomial, in the second interpolation. More exotic machine and deep learning algorithms may further improve the interpolation, reducing the PSF residuals.

In conclusion, our method aims to be a stepping stone towards building better PSF models, reducing the systematic biases induced by the PSF and helping produce more accurate and reliable shear catalogs, which are essential to the understanding of the distribution of dark matter, galaxy evolution, and to constrain cosmological parameters.

\section{acknowledgements}
The Hyper Suprime-Cam (HSC) collaboration includes the astronomical communities of Japan and Taiwan, and Princeton University. The HSC instrumentation and software were developed by the National Astronomical Observatory of Japan (NAOJ), the Kavli Institute for the Physics and Mathematics of the Universe (Kavli IPMU), the University of Tokyo, the High Energy Accelerator Research Organization (KEK), the Academia Sinica Institute for Astronomy and Astrophysics in Taiwan (ASIAA), and Princeton University. Funding was contributed by the FIRST program from the Japanese Cabinet Office, the Ministry of Education, Culture, Sports, Science and Technology (MEXT), the Japan Society for the Promotion of Science (JSPS), Japan Science and Technology Agency (JST), the Toray Science Foundation, NAOJ, Kavli IPMU, KEK, ASIAA, and Princeton University. 

This paper makes use of software developed for the Large Synoptic Survey Telescope. We thank the LSST Project for making their code available as free software at  http://dm.lsst.org

This paper is based [in part] on data collected at the Subaru Telescope and retrieved from the HSC data archive system, which is operated by the Subaru Telescope and Astronomy Data Center (ADC) at National Astronomical Observatory of Japan. Data analysis was in part carried out with the cooperation of Center for Computational Astrophysics (CfCA), National Astronomical Observatory of Japan. The Subaru Telescope is honored and grateful for the opportunity of observing the Universe from Maunakea, which has the cultural, historical and natural significance in Hawaii.

JZ is supported by the National Key Basic Research and Development Program of China (2023YFA1607800, 2023YFA1607802), the NSFC grants (11621303, 11890691, 12073017), and the science research grants from China Manned Space Project (No. CMS-CSST-2021-A01). The computations in this paper were run on the $\pi$ 2.0 cluster supported by the Center of High Performance Computing at Shanghai Jiaotong University, and the Gravity supercomputer of the Astronomy Department, Shanghai Jiaotong University.

\bibliography{main}{}
\bibliographystyle{aasjournal}

\clearpage




\end{document}